\documentclass[aps,prl,twocolumn,showpacs,floatfix,amsmath,superscriptaddress]{revtex4}

\usepackage{bm}
\usepackage{graphicx}
\usepackage{latexsym}
\usepackage{amssymb}
\usepackage{color}

\begin{document}
\bibliographystyle{apsrev}

\title{Spin squeezing in a Rydberg lattice clock}

\author{L.~I.~R.~Gil}
\affiliation{Max Planck Institute for the Physics of Complex Systems, N\"{o}thnitzer Strasse 38, 01187 Dresden, Germany}
\author{R.~Mukherjee}
\affiliation{Max Planck Institute for the Physics of Complex Systems, N\"{o}thnitzer Strasse 38, 01187 Dresden, Germany}
\author{E.~M.~Bridge}
\affiliation{Joint Quantum Centre (JQC) Durham-Newcastle, Department of Physics, Durham University, Durham DH1 3LE, UK}
\author{M.~P.~A.~Jones}
\affiliation{Joint Quantum Centre (JQC) Durham-Newcastle, Department of Physics, Durham University, Durham DH1 3LE, UK}
\author{T.~Pohl}
\affiliation{Max Planck Institute for the Physics of Complex Systems, N\"{o}thnitzer Strasse 38, 01187 Dresden, Germany}

\date{\today}

\begin{abstract}
Squeezed many-body states of atoms  are a valuable resource for high precision frequency metrology and could tremendously boost the performance of atomic lattice clocks. Here, we theoretically demonstrate a viable approach to spin squeezing in lattice clocks via optical dressing of one clock state to a highly excited Rydberg state, generating switchable atomic interactions. For realistic experimental parameters, this is shown to generate over $10$~dB of squeezing in a few microseconds interaction time without affecting the subsequent clock interrogation.
\end{abstract}

\maketitle
The precise measurement of frequency in atomic systems has important applications both in fundamental science, such as tests of relativity \cite{blatt08,chou2010a} and searches for physics beyond the standard model \cite{orzel2012}, and in technologies such as satellite navigation \cite{satt}.
Clocks based on optical transitions have begun to surpass the performance of microwave standards, currently used to define the second \cite{margolis09}. In fact, comparisons between optical clocks based on single ions \cite{chou10} or ensembles of neutral atoms  \cite{nicholson2012,hinkley2013}  are now the most precise measurements ever made. The ensemble approach is spearheaded by Sr and Yb optical lattice clocks, where a fractional frequency instability of  $10^{-17}$ can be obtained in less than one hour \cite{nicholson2012,hinkley2013}. The stability of lattice clocks is now close to the limit imposed  by the quantum projection noise associated with measurements on independent atoms \cite{nicholson2012}. 

Squeezing, or quantum correlations between the atoms, can be used to beat this limit and improve the signal-to-noise ratio \cite{giovannetti04,Ma11}, as was proposed \cite{wineland94,bollinger96} and demonstrated \cite{meyer01,Leibfried2004} in the context of ion traps. Recent experiments have broken the projection noise limit on microwave clock transitions \cite{Louchet-Chauvet2010,Leroux2010}. However, as the state-of-the-art moves towards optical standards, an outstanding challenge is to find an effective method \cite{hald99,kuzmich00,dunn04,meiser08,esteve08,lodewyck09,weinstein10,gross10,torre13,olmos13,rico13} for generating squeezed states in lattice clocks.

In this letter we describe an approach to generating squeezed states of large numbers of atoms in optical lattices of various geometries by exploiting the strong interaction between highly excited atoms. The strong van der Waals interactions between Rydberg atoms provide a
route to creating multi-partite entangled states via the so-called dipole blockade \cite{Lukin2001}, including states suitable for quantum-enhanced measurements  \cite{Bouchoule2002,moller08,saffman09,Mukherjee2011,Opatrny2012}. 
\begin{figure}[t!]
\includegraphics[width = 0.99\columnwidth]{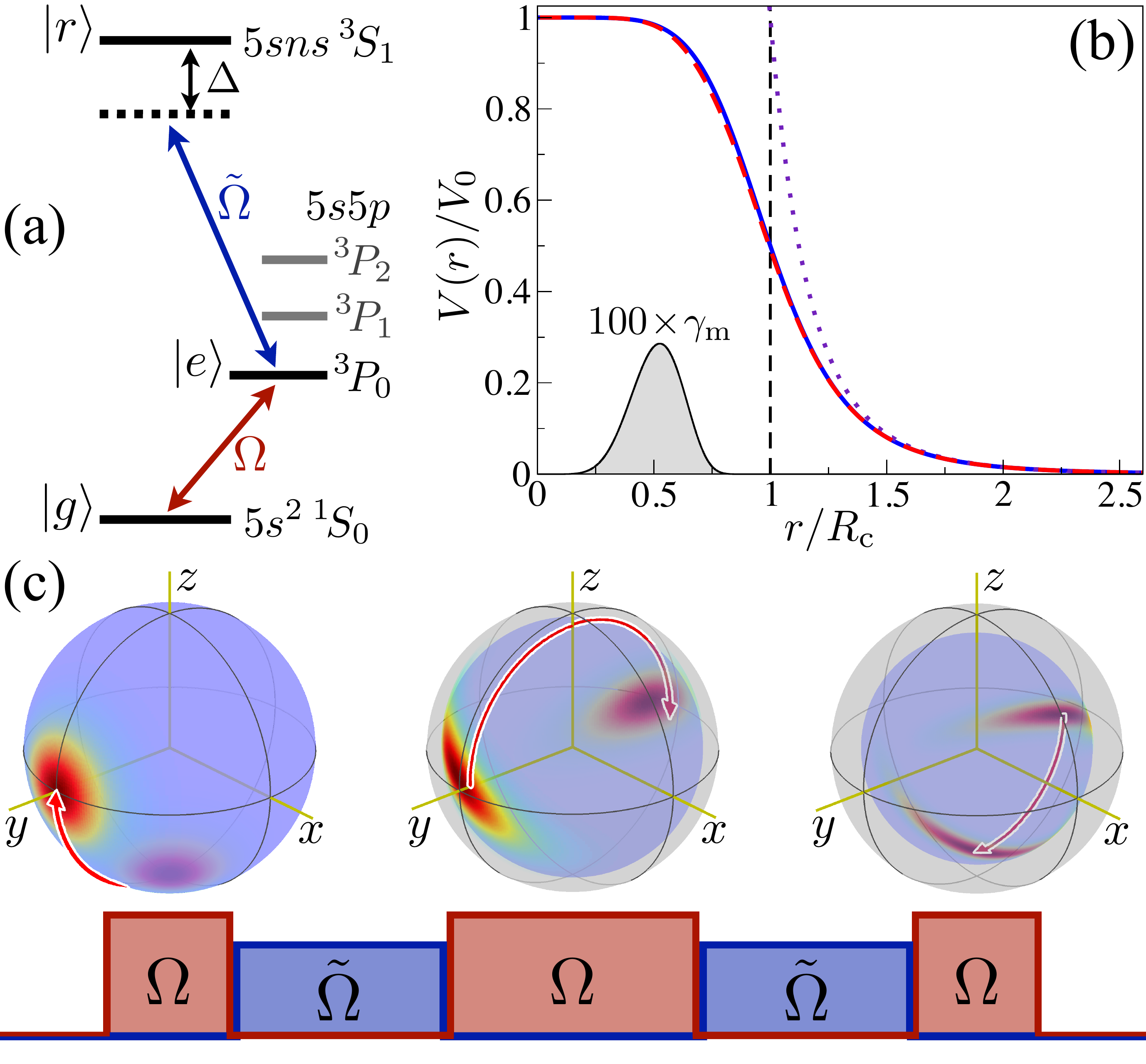}
\caption{\label{fig1} (colour online) (a) Energy level diagram, labelled for the specific example of strontium atoms. Transitions between the clock states $|g\rangle$ and $|e\rangle$ are laser-driven with Rabi frequency $\Omega$. A second laser off-resonantly couples $|e\rangle$ to a high-lying Rydberg state $|r\rangle$ with Rabi frequency $\tilde{\Omega}$. For a large laser detuning $\Delta\gg \tilde{\Omega}$, the system reduces to effective two-level atoms with binary interactions shown by the solid curve in (b). The potential resembles the van der Waals interaction $\sim1/r^6$ (dotted curve) at large separation $r$, but saturates below a critical distance $R_{\rm c}$ (vertical dashed line). In a lattice, the interaction potential is modified slightly (dashed curve), and trap  loss with a small rate $\gamma_{\rm m}$ (grey area) appears. (c) Spin-echo type squeezing protocol, consisting of linear spin rotations around the $x$-axis and nonlinear rotations around the $z$-axis, driven by the two laser fields. The resulting evolution of the total spin is illustrated on a generalized Bloch sphere. For clock operation, this spin-echo sequence is followed by a conventional Rabi or Ramsey scheme.}
\end{figure}
Here, we consider an off-resonant coupling to the Rydberg state, which introduces a switchable, long-range interaction between the atoms in the lattice clock. We show that for realistic experimental parameters, considerable squeezing can be produced within a few microseconds interaction time. The scheme requires only one additional laser, that is switched off prior to clock operation, thereby eliminating additional systematic errors.

An optical lattice clock consists of an ensemble of $N$ atoms trapped in a $d$-dimensional optical lattice with lattice constant $a=\lambda/2$. The clock operates on an inter-combination transition between the singlet ground state ($|g\rangle$) and a long-lived excited triplet state ($|e\rangle$), as shown in Fig.~\ref{fig1}a. The optical lattice is tuned to a magic wavelength, $\lambda$, where the relative Stark shifts of the two states cancel to a very high accuracy \cite{Katori2011}. 
The clock transition is driven with a coupling strength $\Omega$ (see Fig.~\ref{fig1}) for high-precision measurement of the transition energy through Rabi or Ramsey interrogation schemes \cite{westergaard2010}. An ensemble of $N$ such two-level atoms is equivalent to a collection of effective spins described by the spin operators $\hat{\sigma}_{x}^{(i)}=(|g_i\rangle\langle e_i|+|e_i\rangle \langle g_i|)/2$, $\hat{\sigma}_{y}^{(i)}=i(|g_i\rangle\langle e_i|-|e_i\rangle \langle g_i|)/2$ and $\hat{\sigma}_{z}^{(i)}=(|e_i\rangle\langle e_i|-|g_i\rangle \langle g_i|)/2$. One can characterize the many-body states of this system via the total spin $\hat{\bf J}=\sum_i\hat{\bm \sigma}^{(i)}$ that permits convenient visualization of the system dynamics on a generalized Bloch sphere (Fig.~\ref{fig1}c). The precision of any frequency measurement is fundamentally limited by the uncertainty relation $\Delta{\hat{J}_{\perp,1}} \Delta{\hat{J}_{\perp,2}}\geq |\langle \hat{\bf{J}} \rangle|/2$ that relates the variances $\Delta\hat{J}_{\perp,1(2)}$ along two orthogonal directions perpendicular to the mean spin $\langle \hat{\bf{J}} \rangle$ \cite{Robertson}.

For uncorrelated atoms, forming an $N$-particle coherent spin state, the variances $\hat{J}_{\perp,1}=\hat{J}_{\perp,2}=\sqrt{|\langle\hat{\bf J}\rangle|/2}$ prescribe a minimum uncertainty circle limited by quantum projection noise in any direction. The creation of quantum correlations between the atoms permits to reduce the uncertainty $\Delta J_{\perp,{\rm min}}$ along one direction at the expense of the other, leading to a squeezed uncertainty ellipse. The degree of squeezing can be quantified by the squeezing parameter 
\begin{eqnarray}\label{xsi}
\xi^2=\frac{N(\Delta J_{\perp,{\rm min}})^2}{\langle \vec{J} \rangle^2}
\end{eqnarray}
introduced by Wineland \emph{et al.}~\cite{wineland94}, which quantifies directly the gain in precision in a Ramsey spectroscopy measurement relative to an uncorrelated coherent spin state. 

Since atoms are initially uncorrelated, squeezing requires state-dependent atomic interactions \cite{Kitagawa1993,gross12}. The interaction between atoms in the clock states ($|g\rangle$ and $|e\rangle$) is extremely small, although its effect can be detected in lattice clocks \cite{Lemke2011,Swallows2011}. In contrast, the van-der-Waals interaction, $C_6/r^6$ between Rydberg atoms at a distance $r$ takes on enormous values and thereby provides a natural resource for creating correlated states of neutral atoms \cite{wilk10,zhang10,schwarz11,schauss12}. For a strontium lattice clock operating at the magic wavelength of $\lambda=813$nm, the nearest neighbor interaction $C_6/a^6$ between two $|5s50s^{\:3\!}S_1\rangle$-Rydberg state atoms exceeds $10$~GHz \cite{Vaillant2012}. 

These large interaction shifts can be exploited by extending the lattice clock by one additional laser. It couples the long-lived triplet $^{3\!}P_0$ state $|e\rangle$ to a Rydberg state $|r\rangle$  via single photon excitation with a Rabi frequency $\tilde{\Omega}$ (see Fig.~\ref{fig1}). Highly excited atomic states with principal quantum numbers $n=50\ldots100$ have sizable lifetimes on the order of $\tau_{\rm r}\sim10^2~\mu$s \cite{rydlife}, which is a crucial feature of Rydberg atoms for their application in quantum information processing \cite{saff10}. In the present situation, however, the production of large-$N$ squeezed states requires even longer coherence times. Therefore we consider a far off-resonant coupling with detuning $\Delta\gg\tilde{\Omega}$. In this limit the system can be described in terms of effective two-level atoms composed of the unperturbed ground state $|g\rangle$ and a new Rydberg-dressed excited clock state, $|\tilde{e}\rangle\sim|e\rangle-\varepsilon|r\rangle$. Since only a small fraction $\varepsilon=\tilde{\Omega}/(2\Delta)$ of $|r\rangle$ is admixed, the lifetime $\tau_{\rm r}/\varepsilon^2$ of $|\tilde{e}\rangle$ is greatly enhanced. 

Most importantly, the laser coupling induces a light shift $\Delta E_{\rm e}$ of the clock states $|e_i\rangle$ \cite{Bouchoule2002} that, due to the Rydberg-Rydberg atom interaction, is correlated with the positions ${\bf r}_i$ of atoms in $|e\rangle$. From perturbation theory up to fourth order in $\varepsilon$ \cite{Henkel2010},
\begin{equation}
\Delta E_{\rm e}=\sum_i \delta_{\rm e}|e_i\rangle\langle e_i|+\sum_{i<j}V(|{\bf r}_{i}-{\bf r}_{j}|)|e_ie_j\rangle\langle e_ie_j|
\end{equation}
where $\delta_{\rm e}=-\Delta(1-\sqrt{1+4\varepsilon^2})/2$ is the single atom light shift and 
\begin{equation}\label{eq3}
V({\bf r}_i,{\bf r}_j)=V_{ij} = V_0\frac{R^6_c}{|{\bf r}_i-{\bf r}_j|^6 + R_c^6}, 
\end{equation}
corresponds to an effective two-body interaction \cite{Henkel2010,Pupillo2010,Cinti2010,Maucher2011,keating12}. Fig.~\ref{fig1}b illustrates the characteristic  shape of the potential. At large distances, independent dressing of the atoms gives rise to a potential that resembles the original interaction between the Rydberg atoms but with a reduced van der Waals coefficient $\varepsilon^4C_6$. However, below the critical distance $R_{\rm c}=\left|\frac{C_6}{2\hbar\Delta}\right|^{1/6}$ \cite{Henkel2010,Cinti2010,Maucher2011} simultaneous dressing of both atoms is blocked by the interactions such that the effective potential approaches a constant value $V_0 = (\frac{\tilde{\Omega}}{2\Delta})^3 \hbar\tilde{\Omega}$ given by the difference in the light shift of independent and fully blocked atoms \cite{Bouchoule2002}.

Adopting the above spin notation and using this dressed-state picture one obtains the following long-range interacting Ising-type Hamiltonian 
\begin{equation}\label{SpinHamiltonian}
{\cal H}=\hbar \Omega ~\hat{J}_x+ \sum_{i<j}^N V_{ij} \hat{\sigma}_z^{(i)}\hat{\sigma}_z^{(j)}+ \sum_{i}\delta_i \hat{\sigma}_z^{(i)}
\end{equation}
for the effective spins in a transverse field of strength $\hbar\Omega$ and an inhomogeneous longitudinal field $\delta_{i}=\delta_{\rm e}+\frac{1}{2}\sum_{j\neq i}V_{ij}$. Importantly, the transverse ($\hat{H}_{x}=\hbar\Omega\hat{J}_x$) and longitudinal ($\hat{H}_{z}=\sum_{i<j} V_{ij} \hat{\sigma}_z^{(i)}\hat{\sigma}_z^{(j)}+ \sum_{i}\delta_{i} \hat{\sigma}_z^{(i)}$) terms can be turned on and off independently via the intensities of the two laser fields, thus providing great flexibility for the controlled creation of entangled many-body states.

\begin{figure}[t]
\includegraphics[width = 0.99\columnwidth]{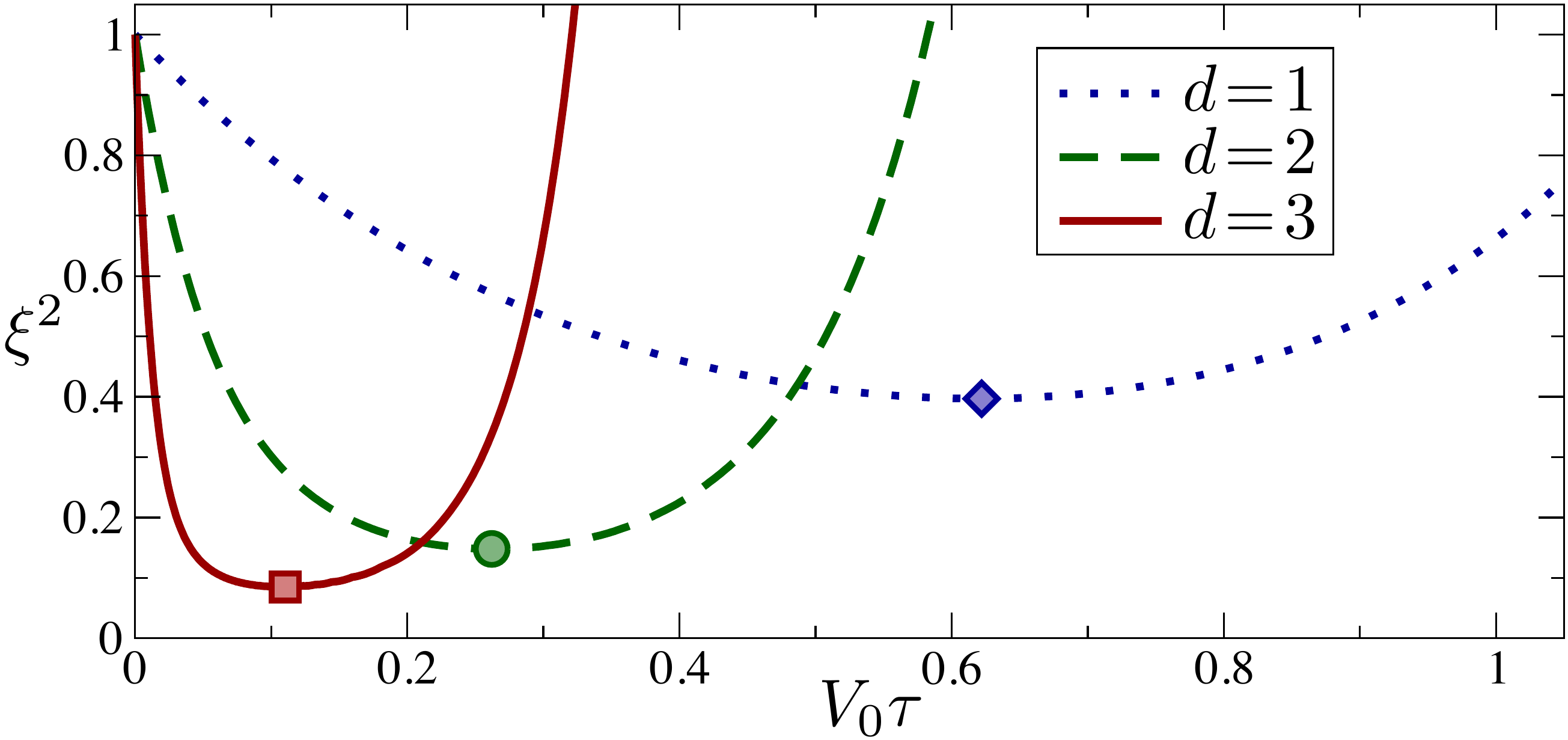}
\caption{\label{fig2}(colour online) Time dependence of the squeezing parameter $\xi^2$ in a $d$-dimensional optical lattice for  $R_c/a=3$. Optimal squeezing is indicated by the symbols.}
\end{figure}

To demonstrate the feasibility of spin-squeezing, we consider here a spin-echo type sequence, illustrated in Fig.~\ref{fig1}c.
Starting from all atoms in the ground state $|g\rangle$ one first applies a $\pi/2$ pulse, with the dressing laser turned off ($\tilde{\Omega}=0$), which rotates the total spin along the $y$-axis (see Fig.~\ref{fig1}c). Subsequent Rydberg dressing, i.e. application of the interaction Hamiltonian $\hat{H}_{z}$, for a time $\tau/2$ then leads to a twisting of the uncertainty ellipse \cite{Kitagawa1993} around the $z$-axis.  In order to eliminate the undesired linear spin-rotation and broadening due to the inhomogeneous detunings $\delta_i$ we subsequently apply a $\pi$-pulse followed by a second dressing phase of duration $\tau/2$ and, finally, another $\pi/2$ that rotates the total spin back along the $z$-axis. The resulting dynamics  can be solved analytically and yields for the final mean spin, $\langle \hat{J}_x\rangle=\langle \hat{J}_y\rangle=0$, 
\begin{equation}\label{eq5}
\langle \hat{J}_z \rangle=-\frac{1}{2} \sum_{i=1}^N \prod_{k \neq i}^N\cos\left(\varphi_{ik} \right)\;,
\end{equation}
where $\varphi_{ij}=V_{ij}\tau/2$. The perpendicular spin component $\hat{J}_{\perp}(\theta)=\cos\left(\theta \right) \hat{J}_x+\sin\left(\theta \right) \hat{J}_y$ at an angle $\theta$ with respect to the $x$-axis has an uncertainty
\begin{eqnarray}\label{eq6}
(\Delta J_{\perp})^2&=&\cos^2\left(\theta \right) \langle \hat{J}_x^2 \rangle+\sin^2\left(\theta \right)  \langle \hat{J}_y^2 \rangle\nonumber \\
&+&\cos\left(\theta \right) \sin\left(\theta \right)\langle \hat{J}_x\hat{J}_y+\hat{J}_y\hat{J}_x \rangle
\end{eqnarray}
where $\langle \hat{J}_y^2 \rangle=N/4$, 
\begin{eqnarray}\label{eq7}
&&\langle \hat{J}_x^2 \rangle=\frac{N}{4}+\frac{1}{4} \sum_{i<j}^N\left[\prod_{k\neq i,j}^N \cos\left(\varphi_{ijk}^{-} \right)-\prod_{k\neq i,j}^N\cos\left(\varphi_{ijk}^+ \right)\right], \nonumber\\
&&\langle \hat{J}_x\hat{J}_y+\hat{J}_y\hat{J}_x \rangle=-\sum_{i<j}^N\sin\left(\varphi_{ij}\right)\prod_{k\neq i,j}^N\cos\left(\varphi_{ik}\right)
\end{eqnarray}
 and $\varphi_{ijk}^{\pm}= \varphi_{ik}\pm\varphi_{jk}$. 
 
In Fig.~\ref{fig2} we show the time evolution of the squeezing parameter $\xi^2$ obtained from eqs.(\ref{eq5})-(\ref{eq7}) upon minimizing $\Delta J_{\perp}$ with respect to $\theta$. For the chosen dressing parameters ($R_{\rm c}/a=3$) the squeezing parameter assumes its minimum value $\xi^2_{\rm min}$ after a short time well below $V_0/\hbar$. Higher dimensional lattices yield stronger squeezing due to a larger number of atoms $N_{\rm c}\sim(R_{\rm c}/a)^d$ within the soft-core radius $R_{\rm c}$.

\begin{figure}[b!]
\includegraphics[width = 0.99\columnwidth]{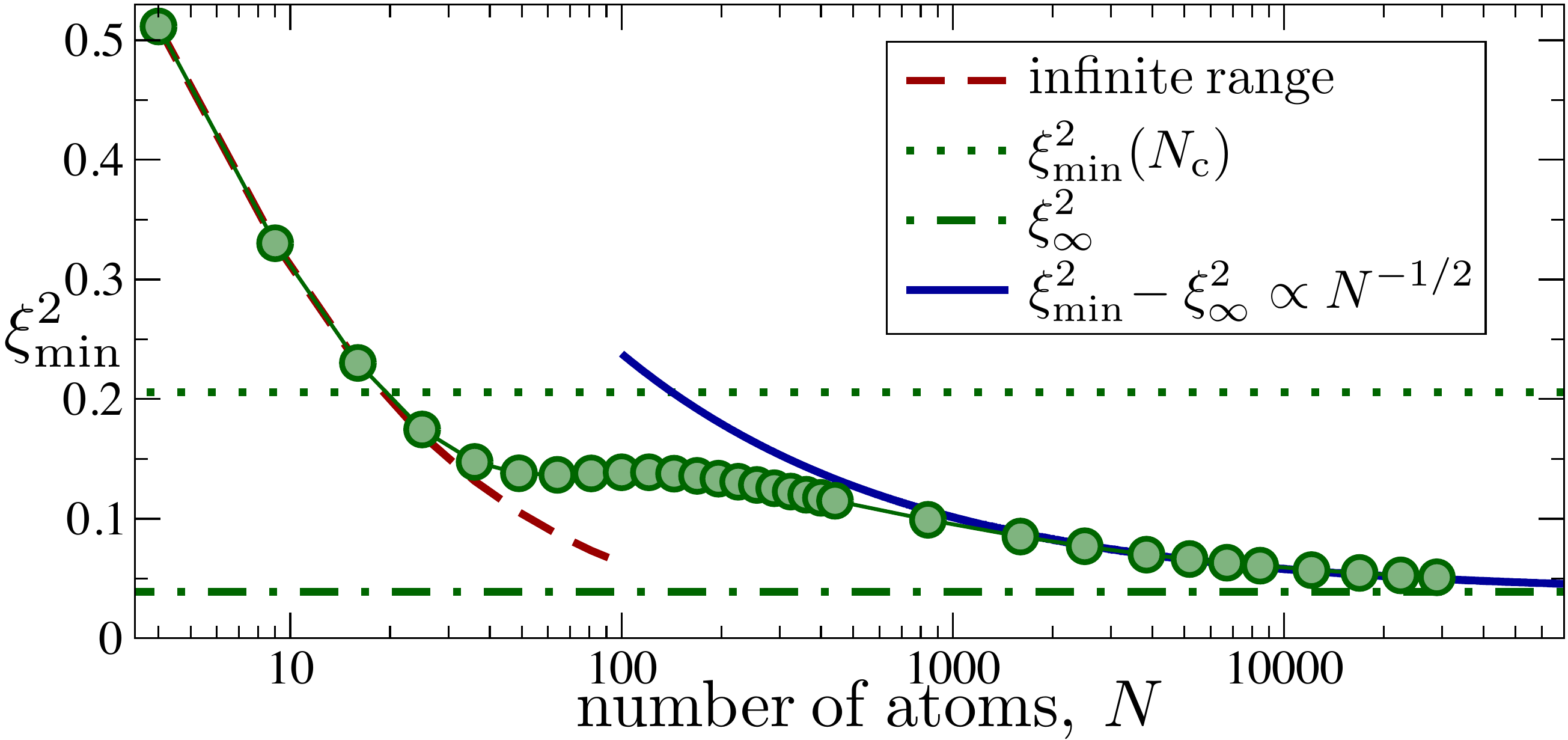}
\caption{\label{fig3}(colour online) Optimal squeezing parameter in a 2D lattice as a function of $N$ for  $R_c/a=5$. The dashed curve shows the result of one-axis twisting by infinite-range interactions \cite{Kitagawa1993}, which agrees with the exact calculations (circles) for small system sizes $N<N_{\rm c}=\pi(R_{\rm c}/2a)^2$ for which all atoms are blocked. However, the squeezing parameter  decreases well below the corresponding value $\xi^2_{\rm min}(N_{\rm c})$ (dotted line). For large $N$, $\xi^2_{\rm min}$ approaches a limiting value $\xi^2_{\infty}$ (dash-dotted line) as $\sim N^{-1/2}$ (thick solid curve).}
\end{figure}

For small fully blockaded systems with $N<N_{\rm c}$ the interaction Hamiltonian reduces to $\hat{H}_z=V_0\hat{J}_z^2/2+\delta \hat{J}_z$, such that the above pulse sequence corresponds to standard one-axis-twisting by infinite-range interactions \cite{Bouchoule2002,Kitagawa1993}. However, as shown in Fig.~\ref{fig3} the squeezing parameter $\xi^2_{\rm min}$ continues to decrease for system sizes $L=N^{1/d}a$ well beyond $R_{\rm c}$, indicating that entanglement between distant spins extends beyond the range of the interactions. For small systems $L<R_{\rm c}$ eqs.(\ref{eq5})-(\ref{eq7}) yield the familiar $\xi^2_{\rm min}\sim N^{-2/3}$ scaling \cite{Kitagawa1993}, as indicated by the dashed line in Fig.~\ref{fig3}. Around $L=R_{\rm c}$, atoms start to explore the finite range of $V_{ij}$ such that the length of the total spin, $\hat{J}^2$, is no longer conserved and the system is driven out of the symmetric Dicke state basis with maximum $\langle\hat{J}^2\rangle=N(N/2+1)/2$. The corresponding decrease of $\langle\hat{J}^2\rangle$ causes an accelerated drop of the final signal $\langle\hat{J}_z\rangle$, leading to a slight increase in $\xi^2_{\rm min}$. However, as the system size is increased further, the continual reduction of $\Delta J_{\perp,{\rm min}}$ compensates for the additional signal loss, such that $\xi^2_{\rm min}$ decreases well below the full-blockade limit \cite{Bouchoule2002} (dotted line in Fig.~\ref{fig3}). 

\begin{figure}
\includegraphics[width = 0.99\columnwidth]{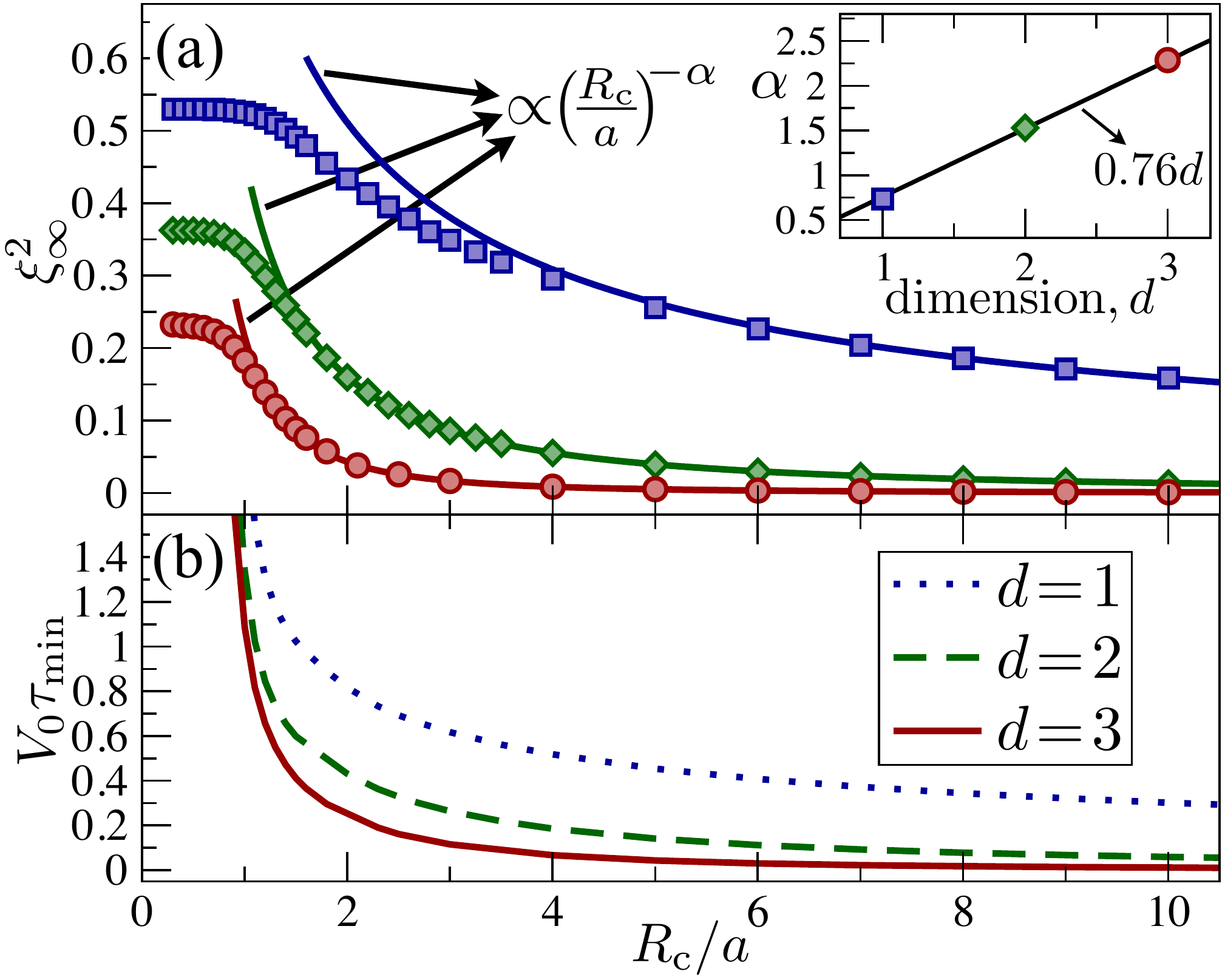}
\caption{\label{fig4}(colour online) (a) Minimal squeezing parameter $\xi^2_{\infty}$ attainable in $d$-dimensional lattices as a function of $R_c/a$. For large $R_{\rm c}$, $\xi^2_{\infty}$ display a power-law decay with an exponent $\alpha\propto d$ (inset).
Panel (b) shows the corresponding interaction time $\tau_{min}$ to realize optimal squeezing.}
\end{figure}

The $N\rightarrow\infty$ limit (dash-dotted line in Fig.~\ref{fig3}) can be calculated efficiently since in this case $\hat{\sigma}_{\alpha}^{(i)}=\hat{\sigma}_{\alpha}^{(j)}$ ($\alpha=x,y,z$). Fig.~\ref{fig4} shows this minimum squeezing parameter $\xi^2_{\infty}$ as a function of the interaction range $R_{\rm c}/a$. For $R_{\rm c}>a$ the squeezing parameter is found to decrease as $\xi^2_{\infty}\sim (R_{\rm c}/a)^{-0.76d}\sim N_{\rm c}^{-0.76}$, i.e. showing a faster drop than small, completely blocked ensembles. For a typical value of $R_{\rm c}/a=5$ we find sizeable squeezing of $\xi^2_{\infty}=0.04$ for 2D and $\xi^2_{\infty}=0.005$ for 3D lattices. Equally important, the time required to obtain optimal squeezing rapidly decreases with $R_{\rm c}/a$. For $R_{\rm c}/a=5$ optimal squeezing is reached after a total dressing time of $\tau=0.14 \hbar/V_0$ for 2D and $\tau=0.04 \hbar/V_0$ for 3D lattices, such that dressing-induced losses can be kept at a low level.

Finally, we consider the specific implementation of this scheme in a Sr lattice clock, including these decoherence mechanisms due to the virtual excitation of Rydberg states. Rydberg excitation of cold Sr atoms has been studied in recent experiments \cite{Millen2010,mcquillen13,lochead13}, and was also considered for blackbody thermometry in Sr lattice clocks \cite{Ovsiannikov2011}. 
In Sr, the $5sns^{\:3\!}S_1$ Rydberg series has nearly isotropic, repulsive van-der-Waals interactions
 \cite{Vaillant2012}, and can be accessed via single-photon excitation from the $5s5p^{\:3\!}P_0$ level with $317$~nm laser light. Tunable, solid-state lasers producing $\sim 0.5$~W of narrowband CW-light have been developed for similar wavelengths \cite{Wilson2011}, which would permit coupling to the $|5s55s^{\:3\!}S_1\rangle$ state with a Rabi frequency of $\tilde{\Omega}/2\pi\approx20$~MHz for a reasonable beam waist of 30 microns. With $\Delta=20\Omega$, these parameters yield a blockade radius of $R_{\rm c}=5a$ and $V_0=2$kHz. According to Fig.~\ref{fig4} this results in a minimum squeezing parameter of $\xi_{\infty}^2=5\times10^{-3}$ within a dressing time of $\tau=20~\mu$s for a 3D lattice clock. This timescale is three orders of magnitude shorter than the lifetime $\tau_{\rm ryd}/\varepsilon^2\approx60$~ms of the Rydberg-dressed clock state, causing simple dephasing of the dressed clock transition. This decreases $\Delta J_{\rm perp}$ by a factor of $\exp(-\varepsilon^2\tau/\tau_{\rm ryd})$ \cite{foss12,foss13} and hence does not affect the attainable squeezing for typical parameters.
  
Operating at the magic wavelength for the two clock states prohibits identical confinement of Rydberg atoms. In fact the Rydberg state polarizability even changes sign \cite{anderson11,anderson12} and yields an inverted lattice potential by a factor $\approx-0.7$ \cite{Mukherjee2011}. However, due to the weak Rydberg-admixture to the excited clock state, its trapping potential is modified only by a negligible fraction of $\sim0.5\varepsilon^2=3\times10^{-4}$, and, hence, does not cause appreciable atomic motion or level shifts. As pointed out recently \cite{li13}, a more significant effect may arise from motional dephasing induced by the strong van der Waals interaction between the virtually excited Rydberg atoms. Adopting the model of \cite{li13}, we have calculated the corresponding loss rate $\gamma_{\rm m}$ and modified interaction potential $\tilde{V}(r)$. As shown in Fig.~\ref{fig1} for a lattice depth of $10$ times the recoil energy $E_{\rm r}$, the resulting potential closely follows eq.(\ref{eq3}) and $\gamma_{\rm m}$ remains below $0.003V_0\approx6$Hz implying a very small trap loss of $\gamma_{\rm m}\tau<10^{-4}$. For tighter lattices with a more typical depth of $100E_{\rm r}$ both effects are further suppressed to a negligible level of $|V(r)-\tilde{V}(r)|<3\times10^{-3}V_0$ and $\gamma_{\rm m}\tau<4\times10^{-6}$. A slight  increase of the excitation level from $n=55$ to $n=75$ doubles the interaction range to $R_{\rm c}=10a$ for which one obtains sizeable spin squeezing of $-30$dB and even smaller losses of $\varepsilon^2\tau/\tau_{\rm ryd}=5\times10^{-5}$ and $\gamma_{\rm m}\tau=2\times10^{-8}$. Importantly, all of these decoherence mechanisms only operate within the interaction time $\tau$, and, hence, do not affect the subsequent measurement stage during which the dressing lasers are switched off. 

In conclusion, we have shown that significant spin squeezing can be obtained in existing optical lattice clocks using only a single additional laser that couples one of the clock states to a Rydberg state.   Due to the strong Rydberg-Rydberg atom interactions, the duration of the presented squeezing protocol is limited only by the available Rabi frequency on the clock transition. Hence, the signal-to-noise ratio can be improved without a significant reduction in duty cycle; an essential prerequisite for practical applications \cite{westergaard11}. Several extensions of the present scheme seem promising.
For example, more involved two-axis twisting schemes \cite{Kitagawa1993} seem possible \cite{Bouchoule2002} and will provide stronger squeezing. Using the present coupling scheme, simultaneous driving by the transverse ($\hat{H}_x$) and longitudinal ($\hat{H}_z$) Hamiltonians \cite{law01,rojo03} or optimized sequences of $\hat{H}_x$ and $\hat{H}_z$ \cite{chen13} could also yield enhanced squeezing.  
More broadly, the availability of long-lived triplet states in two-electron atoms permits $nS$-Rydberg-state dressing via single-photon excitation and, thus, with much higher Rabi frequency and lower decoherence than possible for two-photon dressing of alkaline atoms \cite{Henkel2010}. This opens up a promising route for the exploration of \emph{strong} long-range interactions in Bose-Einstein condensates \cite{cinti12,hsueh12,saccani12,macri13,grusdt13,cinti13} and optical lattices \cite{mattioli13} as well as long-range interacting quantum spin systems, using the setup described in this work.

We thank I. Lesanovsky, K. M{\o}lmer and C. S. Adams for helpful discussions. This work was supported by the EU through the Marie Curie ITN "COHERENCE", and by EPSRC Grant no. EP/J007021/1.


\begin{thebibliography}{100}
\bibitem{blatt08}S. Blatt et al.,  Phys. Rev. Lett. {\bf 100}, 140801 (2008). 
\bibitem{chou2010a}C.~Chou et al., Science, {\bf 329}, 1630 (2010).
\bibitem{orzel2012}C.~Orzel, Physica Scripta {\bf 86}, 068101 (2012).
\bibitem{satt} H. Lichtenegger and E. Wasle \emph{GNSS: Global Navigation Satellite Systems: GPS, GLONASS, Galileo and more}, (Springer, Vienna, 2008).
\bibitem{margolis09}H.~S.~Margolis, Contemporary Physics {\bf 51}, 37 (2009).
\bibitem{chou10}C.~W.~Chou et al., Phys. Rev. Lett. {\bf 104}, 070802 (2010).
\bibitem{nicholson2012}T.~Nicholson \emph{et al.}, Phys. Rev. Lett. {\bf 109}, 230801 (2012).
\bibitem{hinkley2013} N.~Hinkley \emph{et al.}, arXiv:1305.5869.
\bibitem{giovannetti04} V. Giovannetti, S. Lloyd and L. Maccone, Science {\bf 306}, 1330 (2004).
\bibitem{Ma11} J. Ma, X. Wang, C. P. Sun, F. Nori, Phys. Rep. {\bf 509}, 89 (2011).
\bibitem{wineland94}D.~J.~Wineland et al., Phys. Rev. A {\bf 50}, 67 (1994).
\bibitem{bollinger96}J.~J.~Bollinger et al., Phys. Rev. A {\bf 54}, R4649 (1996).
\bibitem{meyer01}V.~Meyer et al., Phys. Rev. Lett. {\bf 86}, 5870 (2001).
\bibitem{Leibfried2004}D.~Leibfried et al., Science {\bf 304}, 1476 (2004).
\bibitem{Louchet-Chauvet2010}A. Louchet-Chauvet et al., New. J. Phys. {\bf 12}, 065032 (2010).
\bibitem{Leroux2010}I.~D.~Leroux, M.~H.~{Schleier-Smith} and V.~Vuleti\'{c}, Phys. Rev. Lett. {\bf 104}, 250801 (2010).
\bibitem{hald99} J. Hald, J.L. S\o{}rensen, C. Schori, E. Polzik, Phys. Rev. Lett. {\bf 83}, 1319 (1999).
\bibitem{kuzmich00} A. Kuzmich, L. Mandel, and N. Bigelow, Phys. Rev. Lett. {\bf 85}, 1594 (2000).
\bibitem{dunn04}J. A. Dunningham and K. Burnett, Phys. Rev. A {\bf 70}, 033601 (2004). 
\bibitem{meiser08}D.~Meiser, J.~Ye and M.~J.~Holland, New J. Phys. {\bf10}, 073014 (2008).
\bibitem{esteve08} J. Est\`eve, C. Gross, A. Weller, S. Giovanazzi1 and M. K. Oberthaler, Nature {\bf 455}, 1216 (2008).
\bibitem{lodewyck09}J.~Lodewyck, P.~G.~Westergaard and P.~Lemonde, Phys. Rev. A {\bf79}, 061401(R) (2009).
\bibitem{weinstein10}J.~D.~Weinstein, K.~Beloy and A.~Derevianko, Phys. Rev. A {\bf 81}, 030302(R) (2010).
\bibitem{gross10} C. Gross, T. Zibold, E. Nicklas, J. Est\`eve, and M. K. Oberthaler, Nature {\bf 464}, 1165 (2010).
\bibitem{torre13} E. G. Dalla Torre, J. Otterbach, E. Demler, V. Vuletic and M. D. Lukin, Phys. Rev. Lett. {\bf 110}, 120402 (2013).
\bibitem{olmos13} B. Olmos, D. Yu, Y. Singh, F. Schreck, K. Bongs, and I. Lesanovsky, Phys. Rev. Lett. {\bf 110}, 143602 (2013).
\bibitem{rico13} L. M. Rico-Gutierrez, T. P. Spiller and J. A. Dunningham, New J. Phys. {\bf 15} (2013).
\bibitem{Lukin2001}M.~D.~Lukin \emph{et al.}, Phys. Rev. Lett. {\bf 87}, 037901 (2001)
\bibitem{Bouchoule2002}I.~Bouchoule and K.~M\o{}lmer, Phys. Rev. A {\bf 65}, 041803(R) (2002).
\bibitem{moller08} D. M\o{}ller,  L.B. Madsen and K. M\o{}lmer, Phys. Rev. Lett., {\bf 100}, 170504 (2008)
\bibitem{saffman09} M. Saffman and K. M\o{}lmer, Phys. Rev. Lett. {\bf 102}, 240502 (2009).
\bibitem{Mukherjee2011}R.~Mukherjee et al., J. Phys. B {\bf 44}, 184010 (2011).
\bibitem{Opatrny2012} T. Opatrn\'y and K. M\o{}lmer, Phys. Rev. A {\bf 86}, 023845 (2012).
\bibitem{Katori2011}H.~Katori, Nature Photonics {\bf 5}, 203 (2011)
\bibitem{westergaard2010}P. ~Westergaard, J. Lodewyck and P. Lemonde, IEEE Trans. Ultrason. Ferroelectr. Freq. Control, {\bf 57}, 623 (2010).
\bibitem{Robertson} H.P.Robertson, Phys. Rev. 34, {\bf 163}164 (1929) 
\bibitem{Kitagawa1993}M.~Kitagawa and M.~Ueda, Phys. Rev. A {\bf 47}, 5138 (1993).
\bibitem{gross12} C. Gross, J. Phys. B {\bf 45}, 103001 (2012).
\bibitem{Swallows2011}M.~D.~Swallows \emph{et al.}, Science {\bf 331}, 1043 (2011).
\bibitem{Lemke2011}N.~D.~Lemke \emph{et al.}, Phys. Rev. Lett. {\bf 107}, 103902 (2011)
\bibitem{wilk10} T. Wilk {\it et al.}, Phys. Rev. Lett. {\bf 104}, 010502 (2010).
\bibitem{zhang10} X. L. Zhang, L. Isenhower, A. T. Gill, T. G. Walker, and M. Saffman, Phys. Rev. A {\bf 82}, 030306 (2010).
\bibitem{schwarz11} A. Schwarzkopf, R. E. Sapiro, and G. Raithel, Phys. Rev. Lett. {\bf 107}, 103001 (2011).
\bibitem{schauss12} P. Schau\ss ~{\it et al.}, Nature {\bf 491}, 87 (2012).
\bibitem{Vaillant2012}C.~L.~Vaillant, M.~P.~A.~Jones and R.~M.~Potvliege, J. Phys. B {\bf 45}, 135004 (2012).
\bibitem{rydlife} S.~Kunze et. al. Z. Phys. D {\bf 27}, 111 (1993); W.~Gornik, Z. Phys A {\bf 283}, 231 (1977).
\bibitem{saff10} M. Saffman, T. Walker and K. M\o{}lmer, Rev. Mod. Phys. {\bf 82}, 2313 (2010).
\bibitem{Henkel2010} N. Henkel, R. Nath and T. Pohl, Phys. Rev. Lett. {\bf 104}, 195302 (2010).
\bibitem{Pupillo2010} G. Pupillo {\it et al.}, Phys. Rev. Lett. {\bf 104}, 223002 (2010).
\bibitem{Cinti2010} F. Cinti {\it et al.}, Phys. Rev. Lett. {\bf 105}, 135301 (2010)
\bibitem{Maucher2011} F. Maucher {\it et al.}, Phys. Rev. Lett. {\bf 106}, 170401 (2011).
\bibitem{keating12} T. Keating {\it et al.}, arXiv:1209.4112.
\bibitem{Millen2010}J.~Millen, G.~Lochead and M.~P.~A.~Jones, Phys. Rev. Lett {\bf 105}, 213004 (2010).
\bibitem{mcquillen13} P. McQuillen, X. Zhang, T. Strickler, F. B. Dunning, and T. C. Killian, Phys. Rev. A {\bf 87}, 013407 (2013).
\bibitem{lochead13} G. Lochead, D. Boddy, D. P. Sadler, C. S. Adams, and M. P. A. Jones, Phys. Rev. A {\bf 87}, 053409 (2013).
\bibitem{Ovsiannikov2011}V.~Ovsiannikov, A.~Derevianko and K.~Gibble, Phys. Rev. Lett. {\bf 107}, 093003 (2011).
\bibitem{Wilson2011} A. Wilson et al., Appl. Phys. B {\bf 105}, 741 (2011).
\bibitem{foss12} M. Foss-Feig, K. R. A. Hazzard, J. J. Bollinger and A. M. Rey, Phys. Rev. A {\bf 87}, 042101 (2012).
\bibitem{foss13} M. Foss-Feig, K. R. A. Hazzard, J. J. Bollinger, A. M. Rey and  C. W. Clark, arxiv:13060172. 
\bibitem{anderson11}S. E. Anderson, K. C. Younge, and G. Raithel, Phys. Rev. Lett. {\bf 107}, 263001 (2011).
\bibitem{anderson12}S. E. Anderson and G. Raithel, Phys. Rev. Lett. {\bf 109}, 023001 (2012).
\bibitem{li13}W. Li, C. Ates, and I. Lesanovsky, Phys. Rev. Lett. {\bf 110}, 213005 (2013).
\bibitem{westergaard11}P.~G.~Westergaard et al., Phys. Rev. Lett. {\bf 106}, 210801 (2011).
\bibitem{law01}C. Law, H. Ng, and P. Leung, Phys. Rev. A {\bf 63}, 055601 (2001).
\bibitem{rojo03}A. Rojo, Phys. Rev. A {\bf 68}, 013807 (2003).
\bibitem{chen13}C. Shen and L. Duan, Phys. Rev. A {\bf 87}, 051801(R) (2013).
\bibitem{cinti12} F. Cinti {\it et al.}, Phys. Rev. Lett. {\bf 108}, 265301 (2012).
\bibitem{hsueh12} C.-H. Hsueh, T.-C. Lin, T.-L. Horng, and W. C. Wu, Phys. Rev. A {\bf 86}, 013619 (2012).
\bibitem{saccani12} S. Saccani, S. Moroni, and M. Boninsegni, Phys. Rev. Lett. {\bf 108}, 175301 (2012).
\bibitem{macri13} T. Macr\`i, F. Maucher, F. Cinti, and T. Pohl, Phys. Rev. A {\bf 87}, 061602(R) (2013).
\bibitem{grusdt13}F. Grusdt and M. Fleischhauer, Phys. Rev. A 87, 043628 (2013).
\bibitem{cinti13} F. Cinti, T. Macr\`i, W. Lechner, G. Pupillo, T. Pohl, arXiv:1302.4576.
\bibitem{mattioli13}M. Mattioli, M. Dalmonte, W. Lechner, G. Pupillo, arXiv:1304.3012.
\end{thebibliography}
\end{document}